\documentclass[prl,amsmath,amssymb,showpacs,showkeys,twocolumn]{revtex4}
\usepackage{graphicx}
\usepackage{dcolumn}
\usepackage{bm}
\begin{document}

\title{ Magnon condensation into $Q$-ball in $^3$He-B}

\author{Yu.~M.~Bunkov$^{(a)}$}
\author{G.E.~Volovik$^{(b,c)}$}
\affiliation{$^{(a)}$Institute Neel, CNRS, Grenoble, France\\
$^{(b)}$Low Temperature Laboratory, Helsinki University of
Technology, Finland\\
$^{(c)}$L.D. Landau Institute for Theoretical Physics,
 Moscow, Russia}

\date{\today}
\begin{abstract}
 The  theoretical prediction of $Q$-balls in relativistic  quantum
fields is realized here experimentally   in superfluid $^3$He-B.  The
condensed-matter analogs of relativistic $Q$-balls are responsible
for an extremely long lived signal of magnetic induction --  the
so-called Persistent Signal -- observed in NMR at the lowest temperatures. This
$Q$-ball  is another representative of a state with phase coherent precession
of nuclear spins in $^3$He-B, similar to the well known
Homogeneously Precessing Domain which we interpret as Bose condensation
of spin waves -- magnons. At large charge $Q$, the effect of self-localization is observed. In the language of relativistic quantum fields it is caused by interaction between the charged and neutral fields, where the neutral field  provides the potential for the charged one. In the process of self-localization the charged field modifies locally the neutral field so that the potential well is formed in which the charge
$Q$ is condensed.

\end{abstract}



\maketitle


A $Q$-ball is a non-topological soliton solution  in field theories containing a complex scalar field $\phi$.  $Q$-balls are stabilized due to the conservation of the global $U(1)$ charge $Q$  \cite{Col}: they exist if the energy minimum develops at nonzero $\phi$ at fixed $Q$.
At the quantum level, $Q$-ball is formed due to suitable attractive interaction
that binds the quanta of $\phi$-field into a large compact object. In some modern SUSY scenarios $Q$-balls are considered as a  heavy particle-like objects, with $Q$ 
being the baryon and/or lepton number.  For many conceivable
alternatives, $Q$-balls may contribute significantly to the dark
matter and baryon contents of the Universe, as described in review
\cite{Enq}. Stable cosmological $Q$-balls can be searched for in existing
and planned experiments \cite{Kus}.

The $Q$-ball is a rather general physical object, which in principle
can be formed  in  condensed matter systems.
In particular, $Q$-balls were suggested  in the atomic Bose-Einstein
condensates \cite{QBallBEC}.
Here we report the observation of $Q$-balls in NMR experiments  in
superfluid $^3$He-B, where the $Q$-balls are formed as special
states of  phase coherent  precession of magnetization. The role
of the $Q$-charge is played by the projection of the total spin of the
system on the axis of magnetic  field, which  is a rather well
conserved quantity at low temperature. At the quantum level, this $Q$-ball is a compact object formed by magnons -- quanta of the corresponding $\phi$-field.

Two types of coherent precession of magnetization have been
observed in
 in superfluid $^3$He-B. The first state known as the  Homogeneously
Precessing Domain (HPD) was discovered in 1984 \cite{HPD}. This
is the bulk state of precessing magnetization which exhibits all the
properties of spin superfluidity and Bose condensation of magnons
(see Reviews \cite{V,BunkovHPDReview}).  These include in particular: spin
supercurrent which transports the magnetization (analog of the mass
current in superfluids and electric supercurrent in
superconductors); spin current Josephson effect and phase-slip
processes at the critical current; and spin current vortex   --  a topological defect  which is the analog of a
quantized vortex in superfluids and of an Abrikosov vortex in
superconductors.

 The HPD loses stability  at low temperatures (below about $0.4~T_c$)
which is now well understood in terms of a parametric instability of
precession towards a decay into pairs of spin waves which is known as the Suhl
instability, see  Ref.\cite{BLV}. In its stead another type  of coherent
precession has been found \cite{PS}. This new state occupies only a
small fraction of the volume of the sample (the signal from such a
state is typically below 1$\%$ of the HPD signal). It can be independently
generated in different parts of the sample, with the signal being
dependent on the position in the container. Such a state lives extremely
long, up to a few minutes, without external pumping \cite{PS, PS2}.
That is why the name Persistent Signal (PS).   We shall show here,
that the PS has the perfect formal analogy with the $Q$-ball.

Both the HPD and PS states can be described in terms of Bose-Einstein
condensation of magnons. The complex order parameter $\Psi$ of the
magnon condensate is related to the precessing spin in the following
way \cite{V}:
\begin{equation}
  \Psi=\sqrt{2S/\hbar}~\sin \frac{\beta}{2}~e^{i\omega t+i\alpha}~,~
S_x+iS_y =S\sin\beta ~e^{i\omega t+i\alpha}. \label{OrderParameter}
\end{equation}
Here ${\bf S}=(S_x,S_y,S_z=S\cos\beta)$ is the vector of spin
density; $\beta$ is the tipping angle of precessing magnetization;
$\omega$ is the precession frequency (in the regime of continuous
NMR, it is the frequency of the applied rf field and it plays the
role of  the chemical potential $\mu=-\omega$ for magnons); $\alpha$
is the phase of precession; $S=\chi H/\gamma$ is  the equilibrium
value of spin density in the applied magnetic  field ${\bf
H}=H\hat{\bf z}$;  $\chi$ is spin suscepti\-bility of liquid
$^3$He-B; and $\gamma$ the gyromagnetic ratio of the $^3$He atom.
The charge $Q$
\begin{equation}
  Q=\int d^3r \vert \Psi\vert^2 =\int d^3r \frac{S-S_z}{\hbar}~,
\label{ChargeQ}
\end{equation}
 is  the number of magnons in the precessing state. It is a
conserved quantity if one neglects the spin-orbit interaction, which is relatively small in $^3$He-B.

The corresponding Gross-Pitaevskii equation is $(\hbar=1)$
  \begin{eqnarray}
(\omega-\omega_L(z)) \Psi= \frac{\delta F}{\delta \Psi^*}~,
  \label{GP}
  \\
 F=\int d^3r\left(\frac{\vert\nabla\Psi\vert^2}{2m_M} +F_D\right).
\label{GL}
\end{eqnarray}
Here $\omega_L(z)=\gamma H(z)$  is the Larmor frequency which
slightly depends on $z$ when the field gradient is applied;   $m_M=\omega_L/2c^2$
is the magnon mass, where $c$ is the spin-wave velocity; and $F_D$ the  spin-orbit (dipole) interaction
averaged over the fast precession, this interaction is typically small so that the precession frequency $\omega\approx\omega_L$. 

The connection to relativistic $Q$-balls becomes clear if one introduces the complex scalar field $\phi=\Psi/\sqrt{\omega}$. Then the gradient energy is $c^2\int d^3r\vert\nabla\phi\vert^2$; the charge is $Q=(1/2i)\int d^3r (\phi^*\partial_t \phi -  \phi\partial_t  \phi^*)$; and solutions which we are looking for have the form
$\phi({\bf r},t)=\phi({\bf r})\exp(i\omega t)$ where the amplitude $\phi({\bf r})\to 0$
as $r\to\infty$.

The  spin-orbit   interaction $F_D$  has a very peculiar
shape in $^3$He-B.   When $\cos\beta >-1/4$ ($\beta< 104^\circ$),
one has
\begin{equation}
  F_D=\chi\Omega_L^2\left[   \frac {4\sin^2(\beta_L/2)}{5S}
\vert\Psi\vert^2- \frac {\sin^4(\beta_L/2)}{S^2} \vert\Psi\vert^4
\right],
     \label{FD}
  \end{equation}
where $\beta_L$ is the polar angle of the vector ${\bf L}$ of
$^3$He-B orbital angular momentum; and $\Omega_L$ is the Leggett
frequency, which characterizes the spin-orbit coupling.  For
$\beta_L=0$ the spin-orbit   interaction becomes identically zero
for $\beta< 104^\circ$, while for larger $\beta> 104^\circ$ one has
\begin{equation}
F_D=\frac{8}{15}\chi\Omega_L^2 \left(
\frac{\vert\Psi\vert^2}{S}-\frac{5}{4}\right)^2 ~~,~ ~\cos\beta
<-1/4~.
     \label{FHPD}
  \end{equation}
 The form of the Ginzburg-Landau functional in Eq.~(\ref{FHPD})
ensures the formation of the Bose condensate when $\omega$ exceeds
$\omega_L$. The Bose condensation of  magnons is the basis for the
HPD:  Larmor precession spontaneously acquires a coherent phase
throughout the whole sample even in an inhomogeneous external
magnetic field. As distinct from the Bose condensates in dilute
gases, the formation of HPD starts with the finite magnitude
$\vert\Psi\vert^2=(5/4)S$, which corresponds to coherent
precession with a tipping angle  equal to the magic Leggett angle, $\beta=
104^\circ$.

For the case, when $\beta_L \neq 0$, the Ginzburg-Landau functional
in Eq.~(\ref{FD}) includes a positive quadratic term and a
negative quartic term, which describes the attractive interaction
between magnons. This form does not support
 the Bose condensation in bulk. However, it is appropriate for  the
formation of the $Q$-ball  in such places in the sample where the potential
produced by the ${\bf L}$-texture
  \begin{equation}
  U({\bf r})=
\frac{4\Omega_L^2}{5\omega_L}\sin^2\frac{\beta_L({\bf r})}{2}~.
\label{Potential2}
\end{equation}
has a minimum.  The precise form of a $Q$-ball depends on the
particular texture and on the position in the container. However, the qualitative behavior of $Q$-balls is generic
and can be illustrated using a simple cylindrically symmetric
distribution of the ${\bf L}$-vector -- the so-called flared-out
texture realized in the geometry of our experimental cell in  the
insert of Fig. 2. Far from the horizontal walls and close to the
axis of the cell the angle $\beta_L$ linearly depends on the
distance $r$ from the axis: $\beta_L({\bf r})\approx \kappa r$  (see
review \cite{SalomaaVolovik}). As a result the potential
(\ref{Potential2}) for magnons is $U(r)\propto \kappa^2r^2$, which
corresponds to the harmonic trap used for confinement of dilute Bose
gases.  The peculiarity of the Ginzburg-Landau functional in
Eq.~(\ref{FD}) is that the quartic term is not a constant but is
$\propto  \kappa^4r^4$. If the $Q$-ball is  formed on the bottom of
the cell (see Fig. 2), the initial potential well is 3-dimensional
and can be approximated as $U(r,z)\propto \kappa^2r^2z^2/(r^2+z^2)$.

To find the $Q$-ball solution, the functional (\ref{GL}) should  be
minimized with fixed number of  charge $Q$ in Eq.~(\ref{ChargeQ}).
Qualitatively this can done using simple dimensional analysis. Let
$r_Q$ and $l_Q$ be the dimensions of the $Q$-ball in radial and in $z$
directions respectively. For the 2D cylindrical $Q$-ball $l_Q=L$ is
the length of the cell, while for the 3D $Q$-ball $l_Q\sim r_Q$. Taking
into account that $Q\sim \vert \Psi\vert^2 r_Q^2l_Q$, one obtains
the energy (\ref{GL}) of $Q$-ball as function of $r_Q$ and $Q$:
 \begin{equation}
F\sim Q\frac{\Omega_L^2}{\omega_L}\left(\frac{\xi_D^2}{r_Q^2}+
\kappa^2r_Q^2-\frac{\kappa^4 Qr_Q^2}{Sl_Q}\right)~,
\label{DimensionalAn}
\end{equation}
where $\xi_D$ is the so-called dipole length characterizing the spin-orbit interaction: in terms of $\xi_D$ magnon mass  $m_M\sim 
\omega_L/( \Omega_L^2\xi_D^2)$. Minimization of Eq.
(\ref{DimensionalAn}) with respect to $r_Q$ gives the equilibrium
radius of the $Q$-ball. While for the 3D $Q$-ball with $l_Q\sim r_Q$ the
equilibrium radius exists for any $Q$, the 2D $Q$-ball is stable
only when $Q<SL/\kappa^2$.

In the regime linear in $Q$ (the regime of spin-waves)   $r_Q\sim \sqrt{\xi_D/\kappa}$ is the characteristic
dimension of a spin wave  localized in the potential well
formed by the flared-out texture. The frequencies of spin wave modes
are
\begin{equation}
 \omega_n-\omega_L= a_n\frac{\Omega_L^2}{\omega_L} \kappa\xi_D ~,
\label{SpinWave}
\end{equation}
where $a_n$ are dimensionless parameters of order unity.  In the 2D
case these are the equidistant levels of the harmonic oscillator,
$a_n\propto (n+1)$, but only even modes $n=0,2,4,...$  are typically
excited in NMR experiments \cite{SalomaaVolovik}. For the 3D case
the spectrum is more complicated, since the potential well is not
harmonic in general.

In the nonlinear regime, one of the energy levels becomes occupied
by magnons. The size of the $Q$-ball  grows with $Q$, and the frequency
starts to decrease.  This behavior is in agreement with the
observation that the Persistent Signal starts to grow from the
linear spin wave localized in the texture \cite{Gren1, Gren2}. During the downward frequency sweep, the spin wave frequency is trapped by
the rf field and after that the amplitude of the signal grows
significantly (Fig. 1). This behavior is similar to that of a
non-linear oscillator with small dissipation. For not very large
 $Q\ll Sl_Q/\kappa^2$, the  frequency of the $Q$-ball  measured
from the frequency of the linear spin wave is
\begin{equation}
 \Delta\omega_Q \sim  -  Q \frac{\Omega_L^2}{\omega_L}\frac{
\kappa^3\xi_D}{Sl_Q} ~. \label{QShift}
\end{equation}
For large $Q$ one must take into account the back reaction of $Q$ to
the texture: the $Q$-ball modifies the potential well as  can be
seen from numerical simulations \cite{Gren3} (Fig.~3).

To compare with the cw NMR experiments, the charge $Q$  must be
expressed in terms of the measurable quantity -- the total
transverse magnetization $M_\perp =\int d^3r S\sin\beta$ in the
persistent signal. For small $\beta$ one has  $M_\perp^2 \sim
r_Q^2l_QSQ \sim \xi_D l_QS/\kappa$, which gives the frequency shift
of the $Q$-ball as a function of the transverse magnetization:
\begin{equation}
 \Delta\omega_Q \sim  -   \frac{\Omega_L^2}{\omega_L}\frac{ \kappa^4
M_\perp^2}{S^2l_Q^2} ~. \label{QShiftM}
\end{equation}
For comparison of the Persistent Signal with the HPD signal we
introduce the total transverse magnetization in the HPD state:
$M_{HPD}\sim SL^3$. Then for the 2D $Q$-ball (with $\kappa^{-1}\sim
l_Q \sim L$) and for the 3D $Q$-balls (with $l_Q= r_Q\sim
\sqrt{\xi_D/\kappa}$ and $\kappa \sim 1/ L$) one obtains
\begin{eqnarray}
   \Delta\omega_Q \sim   -
\frac{\Omega_L^2}{\omega_L}\left(\frac{M_\perp}{M_{HPD}}\right)^2~~,~~2D
\label{QShiftMHPD}
\\
   \Delta\omega_Q \sim   -
\frac{\Omega_L^2}{\omega_L}\frac{L}{\xi_D}
\left(\frac{M_\perp}{M_{HPD}}\right)^2~~,~~3D~. \label{QShiftMHPD3}
\end{eqnarray}

\begin{figure}[ht]
\includegraphics[width=0.45\textwidth]{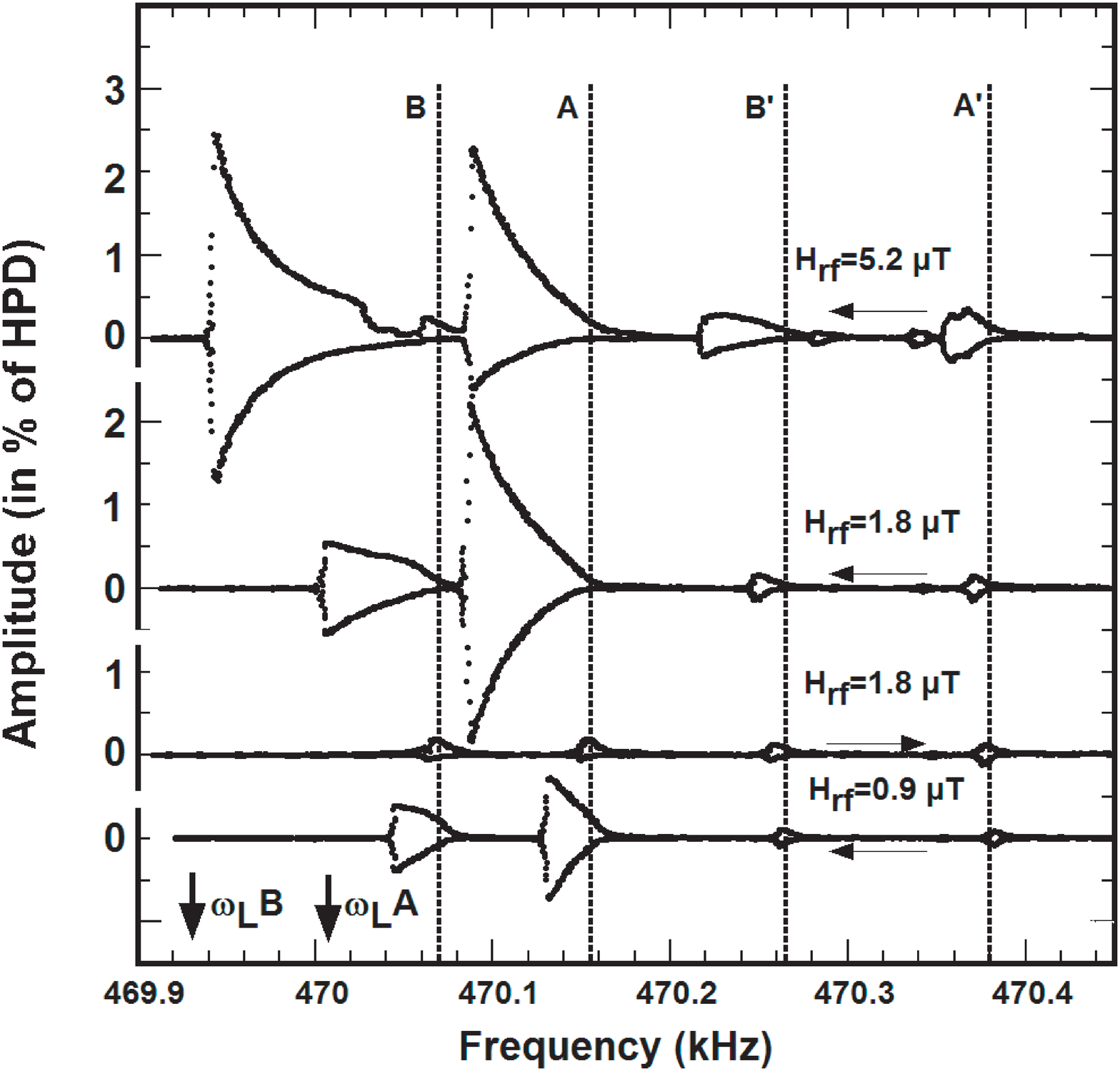}
\vspace{-0.5cm}
 \caption{The cw NMR signals of absorption (negative signals) and full amplitude
of deflected magnetization $M_\perp$ (positive signals) in units of full HPD
signal, as a function of frequency of the rf field measured for
different amplitudes of the rf field, H$_{rf}$, and  directions of
the frequency sweep.  The base line is shifted for each signal.  Larmor frequency
at sites A and B in Fig. 2 is indicated by arrows. At
the sweep down, the spin wave frequency is captured by the rf field
and the amplitude of the signal grows.}
 \label{Fig1}
\end{figure}

\begin{figure}[ht]
\includegraphics[width=0.4\textwidth]{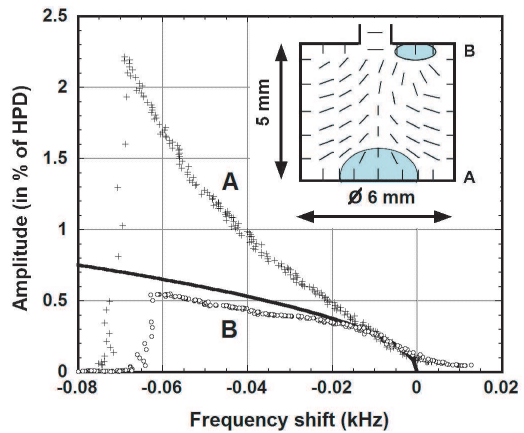}
\vspace{-0.2cm}
 \caption{The amplitude of persistent signal from the $Q$ balls
formed at sites A and B at 1.8 $\mu$T excitation, normalized to the full
HPD signal. The solid curve is the theoretical estimation in
Eq.(\ref{QShiftMHPD3}) with $L/\xi_D\sim 10^2$.}
 \label{Fig2}
\end{figure}

Fig.1 shows the formation of $Q$-balls generated by several spin
wave modes. The signals were recorded at a temperature $T=0.22 T_c$,
pressure 29 bar and different amplitudes of the rf field (for
details, see \cite{Gren1}). By applying different gradients of
magnetic field we are able to localize the positions of the
$Q$-balls. Signals A and A' correspond to the first and second
excited spin-wave modes in the potential well at the bottom of the
cell, while the signals B and B' are from the corresponding modes in the potential
well at the top of the cell (see insert in Fig.2). The 
frequency at which each PS is created corresponds to the frequency of a
linear spin wave in Eq.(\ref{SpinWave}). This allows us to estimate
$\kappa\xi_D\sim 3.5~10^{-3}$. For $\xi_D \sim 10^{-3}$ cm, the inverse $\kappa$ is about 1
cm, which is comparable with the dimension of the cell. This
demonstrates that the texture which produces the potential wells for magnons A and B are generated by the shape of the sample.

With the growing $Q$-ball, its frequency decreases approaching the
Larmor frequency asymptotically. It is useful to normalize the
amplitude of the $Q$-ball signal to the maximum HPD signal, which
corresponds to the case, when all the magnetization inside the cell
is homogeneously precessing with the tipping angle $\beta=
104^\circ$. In Fig.~2 the experimental results for the $Q$-balls A
and B are compared with the theoretical estimations in Eqs.
(\ref{QShiftMHPD}) and (\ref{QShiftMHPD3}). The solid line
corresponds to Eq.~(\ref{QShiftMHPD3}) for the 3D $Q$-ball with the
fitting parameter $L/\xi_D\sim 10^2$, which is consistent with the
dimension $L$ of the cell. In contrast, there is a two order of magnitude
disagreement with Eq.~(\ref{QShiftMHPD}) for the 2D $Q$-ball. This
demonstrates that at least in the initial stage of  $Q$-ball
formation, the $Q$-ball behaves as a 3D object.

\begin{figure}[ht]
\includegraphics[width=0.45\textwidth]{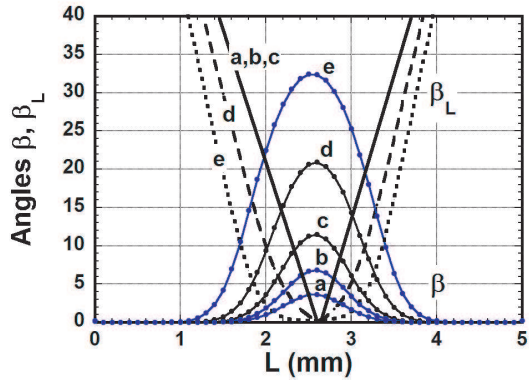}
\vspace{-0.2cm}
 \caption{Numerical simulation of the $Q$-ball growing in
the center of the cell. At small $Q$, when the $\beta$-angle of spin
${\bf S}$ is less than about $10^\circ$, the ${\bf L}$-texture is practically
$Q$-independent  (see $\beta_L(x)$ marked by solid line).
At larger $Q$  the potential well for magnons becomes
wider: the dashed line corresponds to the
${\bf L}$-texture formed by $Q$-ball with $\beta=20^\circ$ in the
middle of the $Q$-ball; and the dotted line --  with $\beta=33^\circ$.
 For details see \cite{Gren3}.}
 \label{Fig3}
\end{figure}

At high excitation one can see a large difference between the two
$Q$-balls, A and B. The more rapid growth of the signal A probably
indicates either the development of the 2D $Q$-ball from the 3D one or the effect
of self-localization shown in Fig. ~3. At high density of magnons
condensed into the $Q$-ball, they influence the ${\bf
L}$-texture  so that the potential well becomes wider and can
incorporate more magnons.  This effect of self-localization  is clearly seen in numerical simulations of interacting ${\bf
L}$ and ${\bf S}$ fields (see Fig. 3 for a one-dimensional $Q$-ball).
At higher excitation both $Q$-balls lose stability.

We discussed here the $Q$-balls formed in potential wells whose
bottom is at $\beta_L=0$. These are not the only possible $Q$-balls.
In the Grenoble experiments \cite{Gren1} the formation of many
different $Q$-balls have been observed. They are generated by spin
waves in the broad range of frequencies. They can
be created in particular by topological defects at the walls of
the cell.

$Q$-balls can be also formed in pulsed NMR measurements. As
distinct from cw NMR, where the $Q$-balls are generated starting
continuously from $Q=0$, in pulsed NMR a $Q$-ball is formed after a
large $Q$ is pumped to the cell. In this case the 3D $Q$-ball is
often formed on the axis of the flared-out texture, away from the
horizontal walls \cite{Lan3}.  This clearly demonstrates the effect
of self-localization: the main part of the pumped charge relaxes but
the rest of $Q$ starts to concentrate at some place on the axis, digging
a potential well there and attracting the charge from other places
of the container. Moreover, this also shows that  $Q$-balls are not necessarily formed at the bottom of the original potential: $Q$-ball may dig the potential well in a different place.  Also it is due to the effect of self-localization that the 2D $Q$-ball is unstable towards the 3D $Q$-ball. 
The $Q$-ball can also be formed with off-resonance excitation \cite{PS2, Gren3}. In this case the effect
of self-localization also plays a crucial role.

In conclusion, the interpretation of an extremely long-lived signal of
magnetic induction in  $^3$He-B in terms of $Q$-ball
formed by quanta of $\phi$-field (magnons) is in  quantitative
agreement with experiments. The $Q$-balls represent a new phase
coherent state of Larmor precession. They emerge at low $T$, when the homogeneous BEC of magnons  (Homogeneously Precessing Domain) becomes unstable. These $Q$-balls are compact
objects which exist due to the conservation of the global $U(1)$ charge (spin projection).
At small $Q$ they are stabilized in the potential well, while
at large $Q$ the effect of self-localization is observed. In terms of relativistic quantum fields the localization is caused by the peculiar interaction between the charged and neutral  fields \cite{FLS}. The neutral field  $\beta_L$ provides the potential for the charged field  $\phi$; the charged field modifies locally the neutral field so that the potential well is formed in which the charge $Q$ is condensed.


We are grateful to H. Godfrin and M. Krusius  for 
discussions. This work  resulted from  collaboration under ULTI project (contract No. RITA-CT-2003-505313) and  between CNRS and Russian Academy of Science (project 19058), and was
supported in part by ESF COSLAB Programme and by RFBR   (grant 06-02-16002-a).

\end{document}